\documentclass{PoS}
\usepackage[title]{appendix}

\title{MAGIC Highlights}

\ShortTitle{MAGIC Highlights}

\author{\speaker{Josep M. Paredes}\thanks{For the MAGIC Collaboration.}\\
        Departament de F\'{\i}sica Qu\`antica i Astrof\'{\i}sica, Institut de Ci\`encies del Cosmos, Universitat de Barcelona, IEEC-UB, Mart\'{\i} i Franqu\`es 1, 08028 Barcelona, Spain\\
        E-mail: \email{jmparedes@ub.edu}}

\abstract{MAGIC has been exploring the sky at Very High Energy gamma-rays (50 GeV -- 50 TeV) since 2004, operating first with a single telescope and from 2009 with two telescopes in stereoscopic mode. MAGIC has carried out a observational program involving fundamental physics and astrophysics topics. In this paper we present some of the most important results obtained by MAGIC, with an special emphasis in the most recent ones.}

\FullConference{Multifrequency Behaviour of High Energy Cosmic Sources - XIII - MULTIF2019\\
		3-8 June 2019\\
		Palermo, Italy}

\begin{document}

\section{Introduction}
\label{intro}
Non-thermal mechanisms are able to accelerate particles up to relativistic energies. There are different scenarios, some of them related to the interaction of winds, jets in AGNs, etc., where strong shocks are produced and particles can gain energy. The relativistic particles accelerated in these environments interact with the ambient matter, the magnetic fields and/or the photon fields to produce gamma-rays. Very High Energy  gamma-rays (VHE, above 100 GeV) can be produced  through  leptonic processes such as Inverse Compton scattering (IC) or through a hadronic process such as proton-proton interactions.

Very high energy cosmic rays or $\gamma$-rays interacting with the molecules of the Earth's atmosphere initiate cascades of particles known as Extensive Air Showers (EAS). The showers produced by primary  $\gamma$-rays are classified as electromagnetic showers whereas the showers produced by hadrons are hadronic showers. Both showers can be distinguished because the hadronic showers produce an asymmetric and wider development compared to electromagnetic showers. The particles produced in such air showers are relativistic and can exceed the speed of light in air, producing the so called Cherenkov radiation \cite{1937PhRv...52..378C}. The Cherenkov light emitted from air showers has a duration of less than 10 nanoseconds and  the peak of the emission is at UV-blue wavelengths (330 nm). This radiation can be detected with Imaging Atmospheric Cherenkov  Telescopes (IACTs) like H.E.S.S., MAGIC or VERITAS.
\section{The MAGIC experiment}
\label{MAGIC}
MAGIC (Major Atmospheric Gamma Imaging Cherenkov) is an international collaboration  of  $\sim$ 160 scientists from 11 countries that operates two IACTs at El Roque de los Muchachos Observatory (2200 m above sea level ) at the island of La Palma, Spain. MAGIC started operating a single telescope with a diameter of 17 m in 2004, and a second telescope with the same diameter was inaugurated in 2009 allowing to operate in stereo mode. The energy range of the stereoscopic system ranges from 50 GeV  to above 50 TeV with an angular resolution of 0.07$^{\circ}$-- 0.14$^{\circ}$ at energies 0.1 -- 1 TeV. The sensitivity is 0.6 \% of the Crab Nebula flux in 50 hours above 300 GeV \cite{2016APh....72...76A} and the field of view is 3.5$^{\circ}$.  The repositioning rate is $\sim$7$^{\circ}$/s. The telescopes performance, in particular the low energy threshold, the high sensitivities within hour timescales and the fast repositioning, makes 
MAGIC an idoneous instrument to explore the VHE sky.

\section{Science Highlights: Galactic}
\label{galactic}
\subsection{Binaries} 
\begin{figure}[htb]
\begin{center}
\includegraphics[width=0.95\textwidth, angle=0]{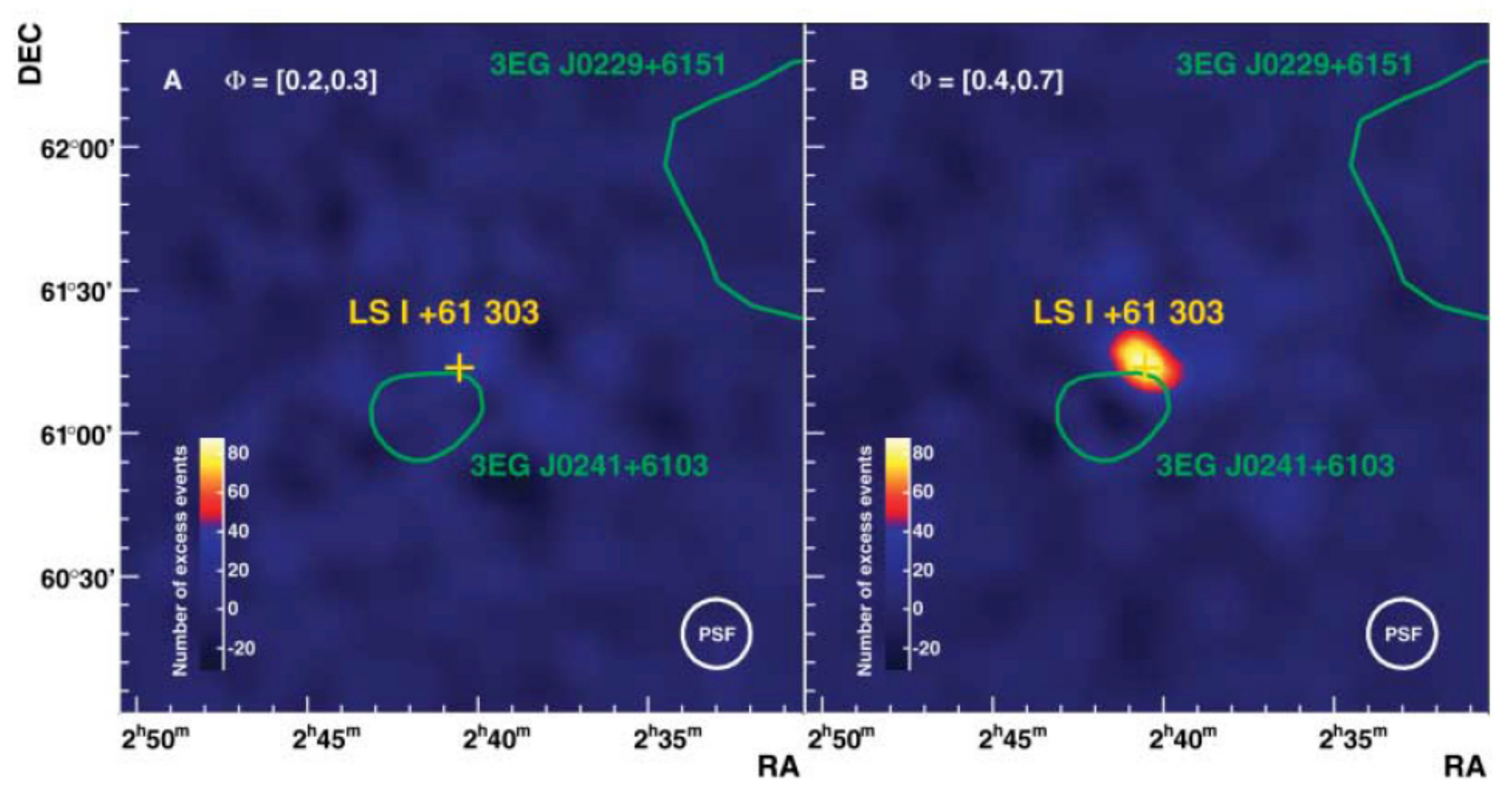}
\caption{Maps of gamma-ray excess above 400 GeV of LS ~I ~+61 303. The left map corresponds to observations around periastron whereas the right map accounts fot observations before the apastron \cite{2006Sci...312.1771A}. \label{fig:lsi}}
\end{center}
\end{figure}

LS ~I ~+61 303 is an X-ray binary system composed of a Be star and a compact object, with an orbital period of 26.5 days and an eccentricity of e = 0.54. Periodic emission coincident with the orbital period has been found in radio, infrared, optical and X-rays, making this source an excellent laboratory to study the different mechanisms taking place in these kind of systems.
Two years after starting its operations, MAGIC detected variable VHE gamma-ray emission from LS ~I ~+61 303 (see figure \ref{fig:lsi})\cite{2006Sci...312.1771A}, suggesting that the emission is periodic and allowing to have a complete view along the full electromagnetic spectrum. The confirmation of the 26.5 days period at VHEs was obtained three years later \cite{2009ApJ...693..303A}. LS ~I ~+61 303 also shows a 4.4 years super-orbital modulation, first suggested by \cite{1987PhDT.......113P} from radio observations. Using VERITAS and MAGIC data, a super-orbital variability at TeV energies was found which is consistent with the radio period \cite{2016A&A...591A..76A}.

Cygnus X-1 and Cygnus X-3 are two binary systems in which the compact object is a well-established black hole in the first case and possibly also in the second one. Both of them have been detected at high-energies by {\it AGILE} \cite{2009Natur.462..620T} and {\it Fermi}-LAT \cite{2009Sci...326.1512F}. At VHEs none of them has been detected and only a hint of emission with MAGIC simultaneously with a hard X-ray flare during a hard state of the source was reported \cite{2007ApJ...665L..51A}.  

\subsection{Pulsars}
\begin{figure}[htb]
\begin{center}
\includegraphics[width=0.9\textwidth]{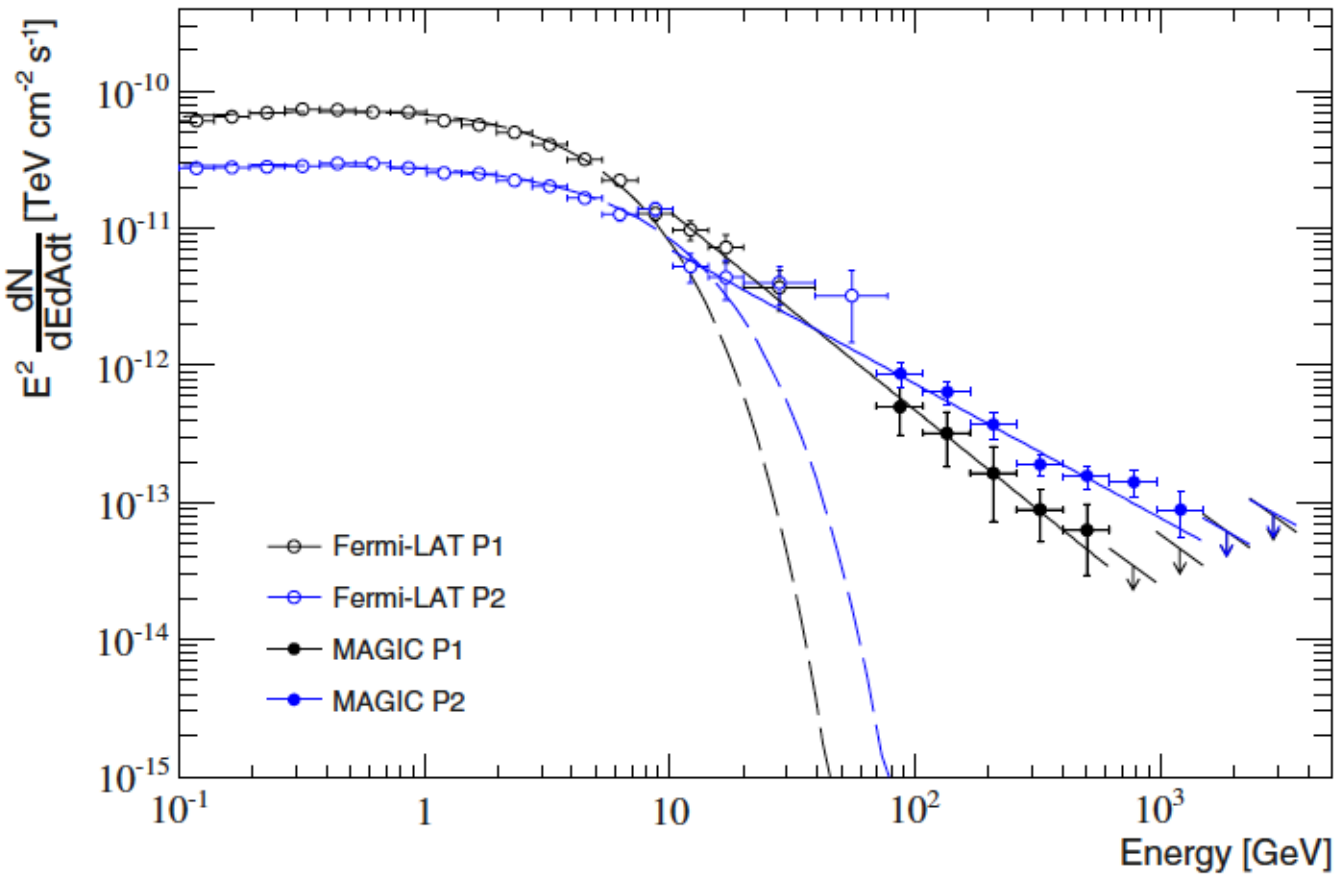}
\caption{Phase-folded Spectral Energy Distribution of the Crab pulses at HE and VHE \cite{2016A&A...585A.133A} \label{fig:crab}.}
\end{center}
\end{figure}

MAGIC detected a pulsed signal from the Crab at $E$ > 25 GeV with a statistical significance of 6.4$\sigma$ \cite{2008Sci...322.1221A}, being the first pulsar seen by a Cherenkov telescope. The pulsed signal occurs at the same spin phases as those observed with EGRET (E > 100 MeV) and simultaneous with optical data taken with the central pixel of the MAGIC camera. The obtained phase-averaged spectrum revealed a high cutoff energy, which indicates that the emission happens far out in the magnetosphere, excluding the polar-cap scenario and challenging the slot-gap scenario. 
At higher energies, VERITAS found pulsations up to energies of 250 GeV \cite{2011Sci...334...69V} and MAGIC up to 400 GeV \cite{2012A&A...540A..69A}. These results motivated a revision of the outer gap model and the generation of new scenarios involving IC scattering of magnetospheric X-rays in presence of a cold ultrarelativistic wind dominated by kinetic energy \cite{2012Natur.482..507A}. More recently, after analyzing more than 300 h of good-quality data obtained during more than seven years of the Crab pulsar observations, MAGIC reported the most energetic pulsed emission ever detected from a pulsar \cite{2016A&A...585A.133A}(see figure \ref{fig:crab}). The pulsed emission, reaching up to 1.5 TeV, challenges  all available models and suggests that the gamma-ray emission is produced in the vicinity of the light cylinder with a mechanism involving IC scattering of low-energy photons.
 
\subsection{Supernova Remnants}
Supernova remnants (SNRs) are thought to accelerate the bulk of the Galactic cosmic rays, and in the case of young SNRs, they could be able to provide PeV cosmic rays. Cassiopeia A (Cas A) is the youngest Galactic SNR, 340 years old, and has been considered an excellent candidate to study particle acceleration processes. The MAGIC telescopes observed Cas A during 158 hours and derived its spectrum between 100 GeV and 10 TeV \cite{2017MNRAS.472.2956A}. The spectrum was completed at lower energies (60 MeV - 500 GeV) by analyzing ~8 yr of {\it Fermi}-LAT data. Both spectra were compatible within errors. The MAGIC data reveal a high-energy cut-off at $\sim$0.01 PeV, challenging the assumption that young SNRs are PeVatrons. The break in the {\it Fermi}-LAT spectrum at $\sim$1 GeV combined with the MAGIC results suggests that the $\gamma$-ray emission is mostly hadronic in origin.

\section{Science Highlights: Extragalactic}
\label{extragalactic}

\subsection{Expanding the TeV universe}
\begin{figure}[htb]
\begin{center}
\includegraphics[width=0.7\textwidth, angle=-90]{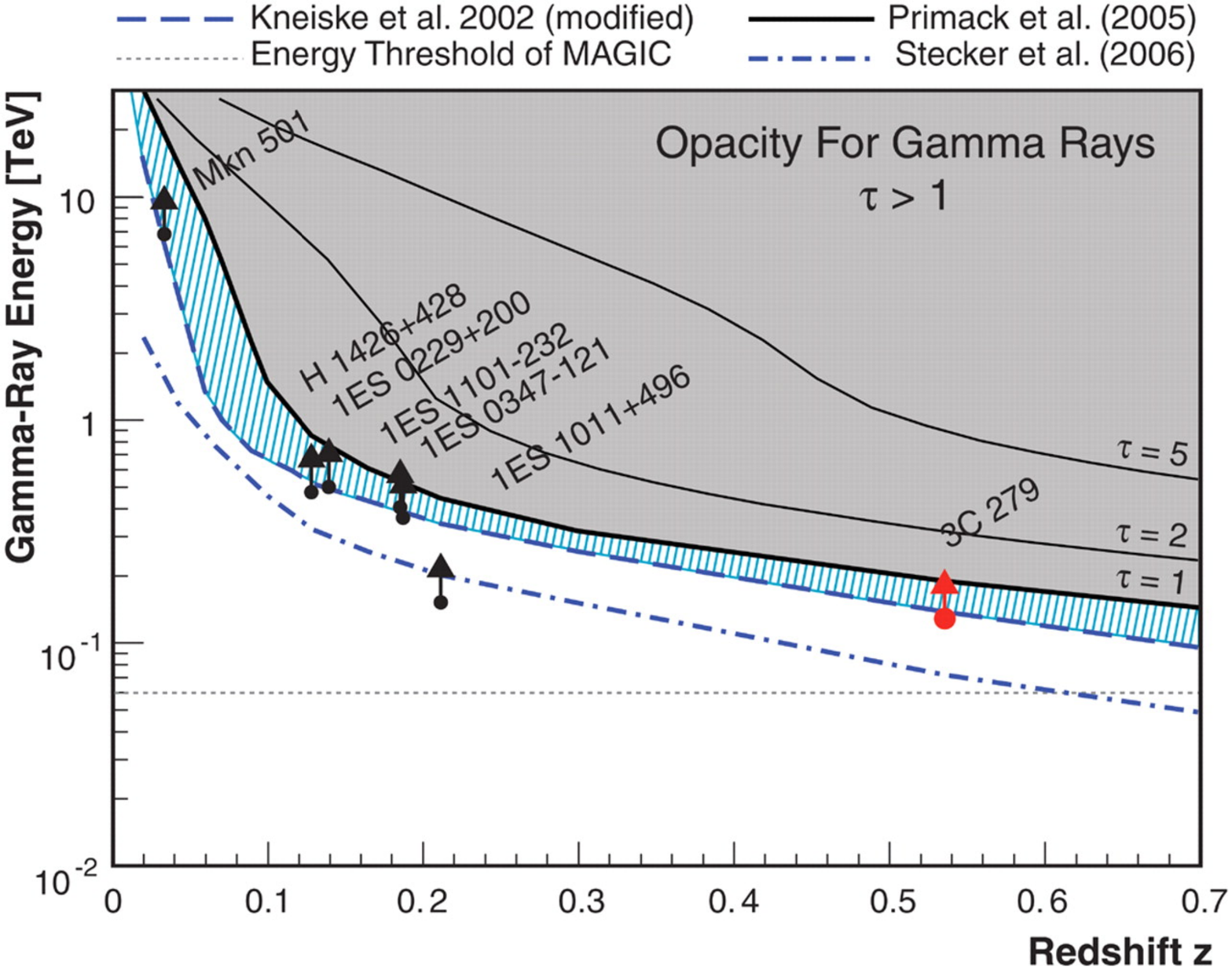}
\caption{The gamma-ray horizon expressed through the energy/redshift relation \cite{2008Sci...320.1752M}. \label{fig:3C279}}
\end{center}
\end{figure}

Highly energetic gamma rays from distant sources can be strongly attenuated after interacting with low-energy photons from the Extragalactic Background Light (EBL) (gamma + gamma --> electron + positron). For z $\sim$ 1 the EBL cut-off is at $\sim$ 100 GeV, at the border of IACT sensitivity. MAGIC, thanks to its low-energy threshold, has been able to detect the farthest objects ever observed in the TeV sky. The first of them was the radio quasar 3C 279, at a redshift of $z$ = 0.536, which allowed to test the transparency of the universe to gamma rays \cite{2008Sci...320.1752M} (see figure \ref{fig:3C279}). Later, with the MAGIC stereoscopic system, VHE emission from the blazar PKS 1441+25 during an outburst was detected for the first time \cite{2015ApJ...815L..23A}. This blazar, with a redshift $z$ = 0.940, is the second most distant known VHE source and probed the EBL at redshifts $\sim$1. Recently, a measurement of the EBL has been carried out using MAGIC and {\it Fermi}-LAT spectra of 12 blazars with  redshift up to 1 \cite{2019MNRAS.486.4233A}.

\subsection{Gravitational lensed $\gamma$-rays}
QSO B0218+357 is a flat spectrum radio quasar located at a redshift of $z$ = 0.944 that is gravitationally lensed by the spiral galaxy B0218+357 located at $z$ = 0.685. Due to the gravitational effect of the intervening galaxy B0218+357, the photons emitted from QSO B0218+357 in the direction to the observer form two paths that reach the observer with a delay of about 11 days \cite{2014ApJ...782L..14C}. In July 2014 QSO B0218+357 experienced a violent flare observed by the {\it Fermi}-LAT and followed by the MAGIC telescopes. This is 
the farthest VHE source detected to date, achieving a flux of $\sim$30 $\%$ of the Crab Nebula at 100 GeV \cite{2016A&A...595A..98A}. The spectral energy distribution of QSO B0218+357 is consistent with current EBL models.

\subsection{Fast variability in AGN}
\begin{figure}[htb]
\begin{center}
\includegraphics[width=0.95\textwidth]{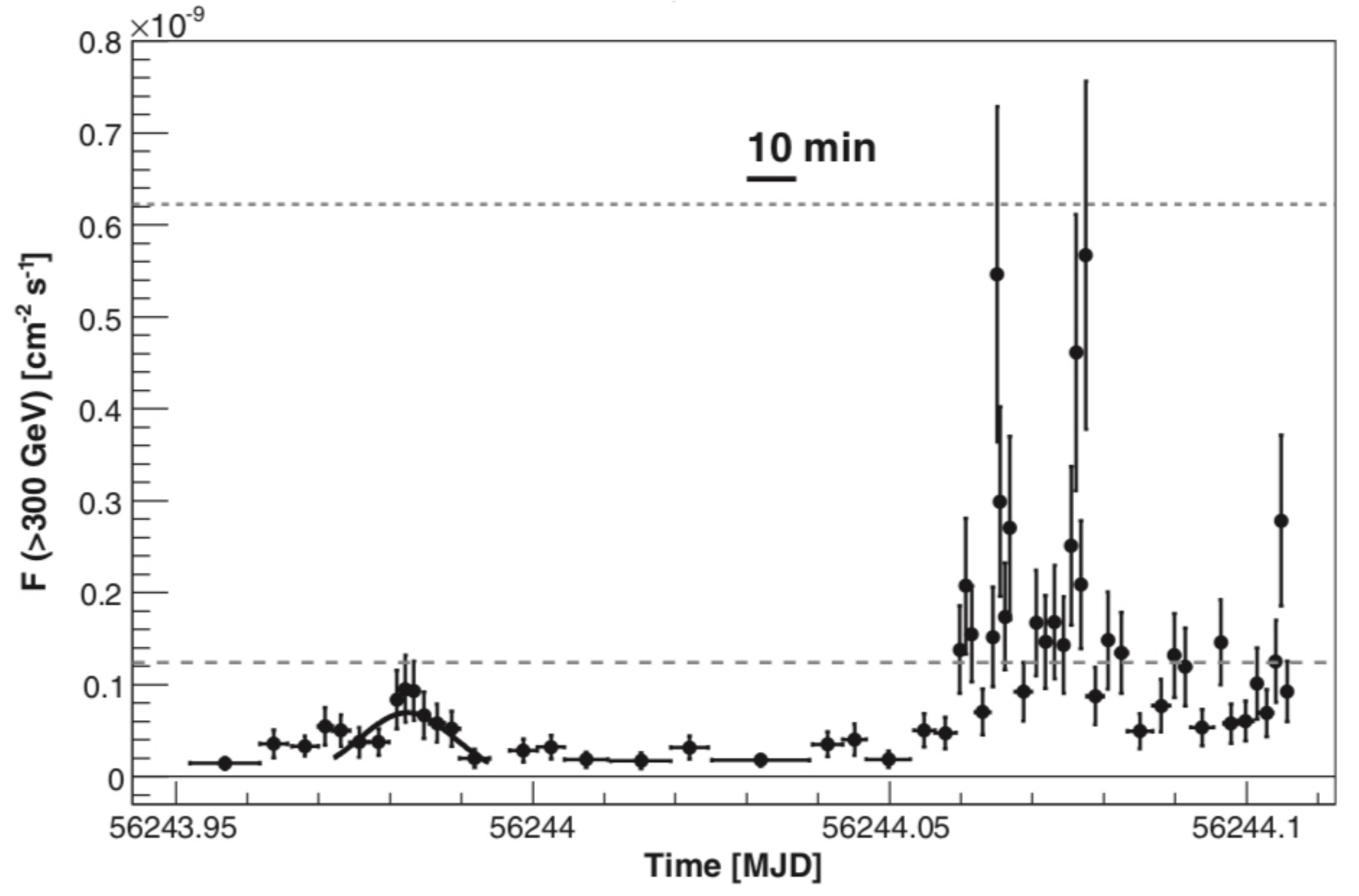}
\caption{Light curve of IC 310 at energies above 300 GeV obtained with the MAGIC telescopes. The two horizontal lines indicate 1 and 5 times the flux level of the Crab Nebula, respectively \cite{2014Sci...346.1080A}. \label{fig:IC310}}
\end{center}
\end{figure}

Supermassive black holes are found in the centers of galaxies where powerful jets are commonly observed at radio wavelengths. MAGIC has detected extremely fast variability from all classes (Radio Galaxies, Flat Spectrum Radio Quasars, BL Lacs).
One way to study the region of formation of the jet is through high resolution radio interferometry observations. However, an alternative method to constrain the size of the emission zone is through variability studies. MAGIC saw an impressive flare of the Radio Galaxy IC 310 in 2014, showing variability with flux doubling in time-scales down to 4.8 minutes \cite{2014Sci...346.1080A} (see figure \ref{fig:IC310}). This fast variability constrains the size of the emitting zone to be smaller than 20$\%$ of the gravitational radius of the black hole. To explain the origin of the substructures smaller than the event horizon, there are different possibilities: mini-jet structures \cite{2010MNRAS.402.1649G}; jet-cloud interactions \cite{2012ApJ...755..170B}; and magnetospheric models \cite{2000A&A...353..473R} where the emission is associated with pulsar-like particle acceleration by the  electric field across a magnetospheric gap at the base of the radio jet. However, it is still unclear whether the emission scenario is close to the central engine or further out. 

\subsection{Dark Matter Searches}
There are different methods to search for Dark Matter (DM). One of these methods is the indirect detection of gamma-rays due to annihilations or decays of Weakly-Interacting Massive Particles (WIMPs) DM particles. Possible targets are:  Clusters of galaxies, the Galactic center and Halo, and dwarf galaxies.

The dwarf spheroidal galaxies are good target candidates for indirect detection of DM with ${\gamma}$-ray observatories. In particular, those who are satellites of the Milky Way are the best candidates because they are relatively close and not much contaminated by ${\gamma}$-rays of astrophysical origin. MAGIC made deep observations of the satellite dwarf galaxy Segue 1. No signals of DM particles in the mass range 100 GeV -- 20 TeV were found \cite{2014JCAP...02..008A}. However, the results provided the most stringent constraints to the annihilation cross-section or decay lifetime obtained from observations of satellite galaxies, for masses above a few hundred GeV.

The MAGIC telescopes have carried out the deepest VHE observational campaign on a Cluster of galaxies, observing the Perseus cluster for about 400 hours in the 2009--2017 period \cite{2018PDU....22...38A}. Although no evidence of a DM signal has been found, it was concluded that DM particles have a decay lifetime longer than $\sim$ $10^{26}$ s in all considered channels, putting the most constraining limits on decaying DM particles from ground-based ${\gamma}$-ray instruments. 

\subsection{MAGIC GRB program}
The design of MAGIC was optimized for working at low energy and with a fast repositioning. These characteristics made MAGIC an instrument with the required performance to study GRBs. Pursuing the goal of detecting a GRB at VHEs, the MAGIC  Collaboration has devoted around 50 hours per year, during many years, to observe GRBs after receiving alerts through  the Gamma-ray Coordinates Network (GCN). Since 2005, 101 GRBs were observed, which means an average of 8--10 GRBs per year. 38 GRBs had the redshift known, and 14 of them a redshift lower than 1.5. Thanks to the MAGIC speed positioning, 22 GRBs were observed with a delay smaller than 100 seconds. 

On January 14, 2019 the X-ray detector BAT, on board the satellite $Swift$, triggered an alert of the GRB 190114C \cite{2019GCN.23688....1G}. The MAGIC telescopes performed a rapid follow-up observation that started about 50s after $Swift$.  The real-time analysis showed a significance of more than 20 $\sigma$ in the first 20 minutes of observations for energies > 300 GeV \cite{2019ATel12390....1M}. This has been the first detection of a GRB at sub-TeV energies.

\section{Science Highlights: Multimessenger}
\label{multi}

\subsection{Gravitational waves follow-up }
An electromagnetic follow-up is necessary to constrain the source's emission of the sources of gravitational waves detected with LIGO/VIRGO. In 2014, a Memorandum of Understanding was signed between LIGO/VIRGO and MAGIC to identify and follow gravitational wave sources. So far, MAGIC has followed up 3 events but it is expected to increase this number in the future.

\subsection{Neutrinos follow-up}
IceCube, located at the South Pole, is the world's largest detector that searches for neutrinos from the most violent astrophysical sources. The production of neutrinos is linked to the production of VHE and UHE cosmic rays. In 2012, MAGIC joined the Gamma-ray Follow-Up (GFU) program. Among the real-time alerts, all the four visible for MAGIC were observed, accounting for more than 30 hours of observation.

\begin{figure}[htb]
\begin{center}
\includegraphics[width=0.95\textwidth]{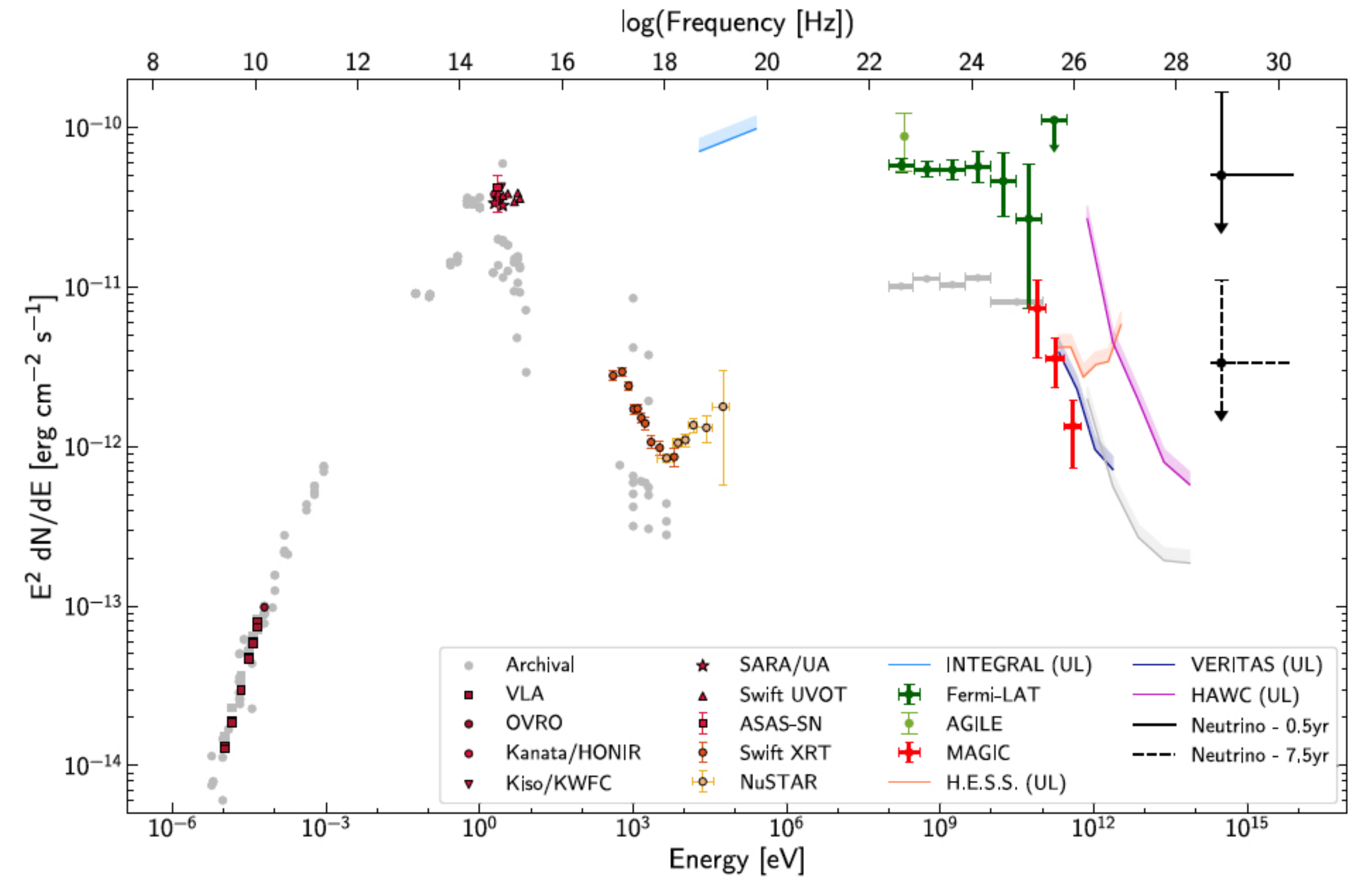}
\caption{Broadband spectral energy distribution for the blazar TXS 0506+056 during the neutrino event recorded in September 28, 2017 \cite{2018Sci...361.1378I}. \label{fig:TXS}}
\end{center}
\end{figure}

On September 22, 2017, the IceCube neutrino observatory detected a high-energy neutrino ($E_{\nu}\sim$ 290 TeV), arriving from a direction consistent with the location of the bright $\gamma$-ray blazar TXS 0506+056 \cite{2018Sci...361.1378I}. There was an immediate follow-up by other instruments. Within the error circle of the event, $Fermi$-LAT found that the blazar TXS 0506+056 was flaring in GeV \cite{2017ATel10791....1T}; also there were  optical and infrared follow-up observations, obtaining a redshift of 0.336 with the 10.4 m Gran Telescopio Canarias \cite{2018ApJ...854L..32P}. MAGIC observed only $\sim$ 2 hours due to bad weather conditions. However, MAGIC resumed observations on 28 September observing the source under good weather conditions and obtaining a 5$\sigma$ detection above 100 GeV after 12 hours of observations \cite{2017ATel10817....1M}. The spectral energy distribution of TXS 0506+056 is shown in figure \ref{fig:TXS}, showing the steep spectrum observed by MAGIC, which is concordant with internal $\gamma$ $\gamma$ absorption above ~100 GeV entailed by photohadronic production of a $\sim$ 290 TeV neutrino.

\section{Summary}
\label{summary}
MAGIC has contributed significantly to the study of the gamma-ray universe, discovering new sources, Crab pulsations and rapid variability of IC 310, extending the EBL studies to $z$ $\sim$1, putting strong constrains on DM, detecting VHE gamma-rays from the blazar  TXS 0506+056 that originated  a high-energy neutrino detected by IceCube, among other results. The most recent result of high impact has been the detection for the first time of a GRB at energies > 300 GeV.  These results have been obtained thanks to the advances in hardware and data analysis development during the last years that have allowed MAGIC to observe a wide variety of sources, with good sensitivity, resolution and fast re-positioning. 

MAGIC now lives its golden age. Both telescopes are working smoothly, obtaining high quality data that will allow the discovery of new sources and advances in the physics  of the known ones. Also, synergies with other facilities operating at lower energies will be strengthened, allowing for a multi-wavelength/multimessenger approach capable to constrain the high energy processes at work in a variety of astronomical sources.

~\\

{\bf Acknowledgements}
This work was supported by the Agencia Estatal de Investigaci\'on grant AYA2016-76012-C3-1-P from the Spanish Ministerio de Econom\'{\i}a y Competitividad (MINECO); by grant MDM-2014-0369 of the ICCUB (Unidad de Excelencia 'Mar\'{\i}a de Maeztu'); and by the Catalan DEC grant 2017 SGR 643, as well as FEDER funds. 


 \bibliographystyle{JHEP} 
 \bibliography{bibliography.bib}

\providecommand{\href}[2]{#2}\begingroup\raggedright\begin{thebibliography}{10}

\bibitem{1937PhRv...52..378C}
P.~A. {{\v{C}}erenkov}, \emph{{Visible Radiation Produced by Electrons Moving
  in a Medium with Velocities Exceeding that of Light}},
  \href{https://doi.org/10.1103/PhysRev.52.378}{\emph{Physical Review}
  {\bfseries 52} (1937) 378}.

\bibitem{2016APh....72...76A}
J.~{Aleksi{\'c}}, S.~{Ansoldi}, L.~{Antonelli}, P.~{Antoranz}, A.~{Babic},
  P.~{Bangale} et~al., \emph{{The major upgrade of the MAGIC telescopes, Part
  II: A performance study using observations of the Crab Nebula}},
  \href{https://doi.org/Astroparticle Physics}{\emph{Astroparticle Physics}
  {\bfseries 72} (2016) 76}
  [\href{https://arxiv.org/abs/astro-ph/1409.5594}{{\ttfamily
  astro-ph/1409.5594}}].

\bibitem{2006Sci...312.1771A}
J.~{Albert}, E.~{Aliu}, H.~{Anderhub}, P.~{Antoranz}, A.~{Armada}, M.~{Asensio}
  et~al., \emph{{Variable Very-High-Energy Gamma-Ray Emission from the
  Microquasar LS I +61 303}},
  \href{https://doi.org/10.1126/science.1128177}{\emph{Science} {\bfseries 312}
  (2006) 1771} [\href{https://arxiv.org/abs/astro-ph/0605549}{{\ttfamily
  astro-ph/0605549}}].

\bibitem{2009ApJ...693..303A}
J.~{Albert}, E.~{Aliu}, H.~{Anderhub}, L.~A. {Antonelli}, P.~{Antoranz},
  M.~{Backes} et~al., \emph{{Periodic Very High Energy
  {\ensuremath{\gamma}}-Ray Emission from LS I +61 303 Observed
  with the MAGIC Telescope}},
  \href{https://doi.org/10.1088/0004-637X/693/1/303}{\emph{\apj} {\bfseries
  693} (2009) 303} [\href{https://arxiv.org/abs/0806.1865}{{\ttfamily
  0806.1865}}].

\bibitem{1987PhDT.......113P}
J.~M. {Paredes}, Ph.D. thesis, University of Barcelona, Sep, 1987.

\bibitem{2016A&A...591A..76A}
M.~L. {Ahnen}, S.~{Ansoldi}, L.~A. {Antonelli}, P.~{Antoranz}, A.~{Babic},
  B.~{Banerjee} et~al., \emph{{Super-orbital variability of LS I
  +61 303 at TeV energies}},
  \href{https://doi.org/10.1051/0004-6361/201527964}{\emph{\aap} {\bfseries
  591} (2016) A76} [\href{https://arxiv.org/abs/1603.06973}{{\ttfamily
  1603.06973}}].

\bibitem{2009Natur.462..620T}
M.~{Tavani}, A.~{Bulgarelli}, G.~{Piano}, S.~{Sabatini}, E.~{Striani},
  Y.~{Evangelista} et~al., \emph{{Extreme particle acceleration in the
  microquasar CygnusX-3}},
  \href{https://doi.org/10.1038/nature08578}{\emph{\nat} {\bfseries 462} (2009)
  620} [\href{https://arxiv.org/abs/0910.5344}{{\ttfamily 0910.5344}}].

\bibitem{2009Sci...326.1512F}
{Fermi LAT Collaboration}, A.~A. {Abdo}, M.~{Ackermann}, M.~{Ajello},
  M.~{Axelsson}, L.~{Baldini} et~al., \emph{{Modulated High-Energy Gamma-Ray
  Emission from the Microquasar Cygnus X-3}},
  \href{https://doi.org/10.1126/science.1182174}{\emph{Science} {\bfseries 326}
  (2009) 1512}.

\bibitem{2007ApJ...665L..51A}
J.~{Albert}, E.~{Aliu}, H.~{Anderhub}, P.~{Antoranz}, A.~{Armada},
  C.~{Baixeras} et~al., \emph{{Very High Energy Gamma-Ray Radiation from the
  Stellar Mass Black Hole Binary Cygnus X-1}},
  \href{https://doi.org/10.1086/521145}{\emph{\apjl} {\bfseries 665} (2007)
  L51} [\href{https://arxiv.org/abs/0706.1505}{{\ttfamily 0706.1505}}].

\bibitem{2008Sci...322.1221A}
E.~{Aliu}, H.~{Anderhub}, L.~A. {Antonelli}, P.~{Antoranz}, M.~{Backes},
  C.~{Baixeras} et~al., \emph{{Observation of Pulsed {\ensuremath{\gamma}}-Rays
  Above 25 GeV from the Crab Pulsar with MAGIC}},
  \href{https://doi.org/10.1126/science.1164718}{\emph{Science} {\bfseries 322}
  (2008) 1221} [\href{https://arxiv.org/abs/0809.2998}{{\ttfamily 0809.2998}}].

\bibitem{2011Sci...334...69V}
{VERITAS Collaboration}, E.~{Aliu}, T.~{Arlen}, T.~{Aune}, M.~{Beilicke},
  W.~{Benbow} et~al., \emph{{Detection of Pulsed Gamma Rays Above 100 GeV from
  the Crab Pulsar}},
  \href{https://doi.org/10.1126/science.1208192}{\emph{Science} {\bfseries 334}
  (2011) 69} [\href{https://arxiv.org/abs/1108.3797}{{\ttfamily 1108.3797}}].

\bibitem{2012A&A...540A..69A}
J.~{Aleksi{\'c}}, E.~A. {Alvarez}, L.~A. {Antonelli}, P.~{Antoranz},
  M.~{Asensio}, M.~{Backes} et~al., \emph{{Phase-resolved energy spectra of the
  Crab pulsar in the range of 50-400 GeV measured with the MAGIC telescopes}},
  \href{https://doi.org/10.1051/0004-6361/201118166}{\emph{\aap} {\bfseries
  540} (2012) A69} [\href{https://arxiv.org/abs/1109.6124}{{\ttfamily
  1109.6124}}].

\bibitem{2012Natur.482..507A}
F.~A. {Aharonian}, S.~V. {Bogovalov} and D.~{Khangulyan}, \emph{{Abrupt
  acceleration of a `cold' ultrarelativistic wind from the Crab pulsar}},
  \href{https://doi.org/10.1038/nature10793}{\emph{\nat} {\bfseries 482} (2012)
  507}.

\bibitem{2016A&A...585A.133A}
S.~{Ansoldi}, L.~A. {Antonelli}, P.~{Antoranz}, A.~{Babic}, P.~{Bangale},
  U.~{Barres de Almeida} et~al., \emph{{Teraelectronvolt pulsed emission from
  the Crab Pulsar detected by MAGIC}},
  \href{https://doi.org/10.1051/0004-6361/201526853}{\emph{\aap} {\bfseries
  585} (2016) A133} [\href{https://arxiv.org/abs/1510.07048}{{\ttfamily
  1510.07048}}].

\bibitem{2017MNRAS.472.2956A}
M.~L. {Ahnen}, S.~{Ansoldi}, L.~A. {Antonelli}, C.~{Arcaro}, A.~{Babi{\'c}},
  B.~{Banerjee} et~al., \emph{{A cut-off in the TeV gamma-ray spectrum of the
  SNR Cassiopeia A}},
  \href{https://doi.org/10.1093/mnras/stx2079}{\emph{\mnras} {\bfseries 472}
  (2017) 2956} [\href{https://arxiv.org/abs/1707.01583}{{\ttfamily
  1707.01583}}].

\bibitem{2008Sci...320.1752M}
{MAGIC Collaboration}, J.~{Albert}, E.~{Aliu}, H.~{Anderhub}, L.~A.
  {Antonelli}, P.~{Antoranz} et~al., \emph{{Very-High-Energy gamma rays from a
  Distant Quasar: How Transparent Is the Universe?}},
  \href{https://doi.org/10.1126/science.1157087}{\emph{Science} {\bfseries 320}
  (2008) 1752} [\href{https://arxiv.org/abs/0807.2822}{{\ttfamily 0807.2822}}].

\bibitem{2015ApJ...815L..23A}
M.~L. {Ahnen}, S.~{Ansoldi}, L.~A. {Antonelli}, P.~{Antoranz}, A.~{Babic},
  B.~{Banerjee} et~al., \emph{{Very High Energy {\ensuremath{\gamma}}-Rays from
  the Universe's Middle Age: Detection of the z = 0.940 Blazar PKS 1441+25 with
  MAGIC}}, \href{https://doi.org/10.1088/2041-8205/815/2/L23}{\emph{\apjl}
  {\bfseries 815} (2015) L23}
  [\href{https://arxiv.org/abs/1512.04435}{{\ttfamily 1512.04435}}].

\bibitem{2019MNRAS.486.4233A}
V.~A. {Acciari}, S.~{Ansoldi}, L.~A. {Antonelli}, A.~{Arbet Engels},
  D.~{Baack}, {Babi{\'c}} et~al., \emph{{Measurement of the extragalactic
  background light using MAGIC and Fermi-LAT gamma-ray observations of blazars
  up to z = 1}}, \href{https://doi.org/10.1093/mnras/stz943}{\emph{\mnras}
  {\bfseries 486} (2019) 4233}
  [\href{https://arxiv.org/abs/1904.00134}{{\ttfamily 1904.00134}}].

\bibitem{2014ApJ...782L..14C}
C.~C. {Cheung}, S.~{Larsson}, J.~D. {Scargle}, M.~A. {Amin}, R.~D. {Blandford},
  D.~{Bulmash} et~al., \emph{{Fermi Large Area Telescope Detection of
  Gravitational Lens Delayed {\ensuremath{\gamma}}-Ray Flares from Blazar
  B0218+357}}, \href{https://doi.org/10.1088/2041-8205/782/2/L14}{\emph{\apjl}
  {\bfseries 782} (2014) L14}
  [\href{https://arxiv.org/abs/1401.0548}{{\ttfamily 1401.0548}}].

\bibitem{2016A&A...595A..98A}
M.~L. {Ahnen}, S.~{Ansoldi}, L.~A. {Antonelli}, P.~{Antoranz}, C.~{Arcaro},
  A.~{Babic} et~al., \emph{{Detection of very high energy gamma-ray emission
  from the gravitationally lensed blazar QSO B0218+357 with the MAGIC
  telescopes}}, \href{https://doi.org/10.1051/0004-6361/201629461}{\emph{\aap}
  {\bfseries 595} (2016) A98}
  [\href{https://arxiv.org/abs/1609.01095}{{\ttfamily 1609.01095}}].

\bibitem{2014Sci...346.1080A}
J.~{Aleksi{\'c}}, S.~{Ansoldi}, L.~A. {Antonelli}, P.~{Antoranz}, A.~{Babic},
  P.~{Bangale} et~al., \emph{{Black hole lightning due to particle acceleration
  at subhorizon scales}},
  \href{https://doi.org/10.1126/science.1256183}{\emph{Science} {\bfseries 346}
  (2014) 1080} [\href{https://arxiv.org/abs/1412.4936}{{\ttfamily 1412.4936}}].

\bibitem{2010MNRAS.402.1649G}
D.~{Giannios}, D.~A. {Uzdensky} and M.~C. {Begelman}, \emph{{Fast TeV
  variability from misaligned minijets in the jet of M87}},
  \href{https://doi.org/10.1111/j.1365-2966.2009.16045.x}{\emph{\mnras}
  {\bfseries 402} (2010) 1649}
  [\href{https://arxiv.org/abs/0907.5005}{{\ttfamily 0907.5005}}].

\bibitem{2012ApJ...755..170B}
M.~V. {Barkov}, V.~{Bosch-Ramon} and F.~A. {Aharonian}, \emph{{Interpretation
  of the Flares of M87 at TeV Energies in the Cloud-Jet Interaction Scenario}},
  \href{https://doi.org/10.1088/0004-637X/755/2/170}{\emph{\apj} {\bfseries
  755} (2012) 170} [\href{https://arxiv.org/abs/1202.5907}{{\ttfamily
  1202.5907}}].

\bibitem{2000A&A...353..473R}
F.~M. {Rieger} and K.~{Mannheim}, \emph{{Particle acceleration by rotating
  magnetospheres in active galactic nuclei}}, {\emph{\aap} {\bfseries 353}
  (2000) 473} [\href{https://arxiv.org/abs/astro-ph/9911082}{{\ttfamily
  astro-ph/9911082}}].

\bibitem{2014JCAP...02..008A}
J.~{Aleksi{\'c}}, S.~{Ansoldi}, L.~A. {Antonelli}, P.~{Antoranz}, A.~{Babic},
  P.~{Bangale} et~al., \emph{{Optimized dark matter searches in deep
  observations of Segue 1 with MAGIC}},
  \href{https://doi.org/10.1088/1475-7516/2014/02/008}{\emph{\jcap} {\bfseries
  2014} (2014) 008} [\href{https://arxiv.org/abs/1312.1535}{{\ttfamily
  1312.1535}}].

\bibitem{2018PDU....22...38A}
V.~A. {Acciari}, S.~{Ansoldi}, L.~A. {Antonelli}, A.~{Arbet Engels},
  C.~{Arcaro}, D.~{Baack} et~al., \emph{{Constraining dark matter lifetime with
  a deep gamma-ray survey of the Perseus galaxy cluster with MAGIC}},
  \href{https://doi.org/10.1016/j.dark.2018.08.002}{\emph{Physics of the Dark
  Universe} {\bfseries 22} (2018) 38}
  [\href{https://arxiv.org/abs/1806.11063}{{\ttfamily 1806.11063}}].

\bibitem{2019GCN.23688....1G}
J.~D. {Gropp}, J.~A. {Kennea}, N.~J. {Klingler}, H.~A. {Krimm}, S.~J.
  {Laporte}, A.~Y. {Lien} et~al., \emph{{GRB 190114C: Swift detection of a very
  bright burst with a bright optical counterpart.}}, {\emph{GRB Coordinates
  Network} {\bfseries 23688} (2019) 1}.

\bibitem{2019ATel12390....1M}
R.~{Mirzoyan}, \emph{{First time detection of a GRB at sub-TeV energies; MAGIC
  detects the GRB 190114C}}, {\emph{The Astronomer's Telegram} {\bfseries
  12390} (2019) 1}.

\bibitem{2018Sci...361.1378I}
{IceCube Collaboration}, M.~G. {Aartsen}, M.~{Ackermann}, J.~{Adams}, J.~A.
  {Aguilar}, M.~{Ahlers} et~al., \emph{{Multimessenger observations of a
  flaring blazar coincident with high-energy neutrino IceCube-170922A}},
  \href{https://doi.org/10.1126/science.aat1378}{\emph{Science} {\bfseries 361}
  (2018) eaat1378} [\href{https://arxiv.org/abs/1807.08816}{{\ttfamily
  1807.08816}}].

\bibitem{2017ATel10791....1T}
Y.~T. {Tanaka}, S.~{Buson} and D.~{Kocevski}, \emph{{Fermi-LAT detection of
  increased gamma-ray activity of TXS 0506+056, located inside the
  IceCube-170922A error region.}}, {\emph{The Astronomer's Telegram} {\bfseries
  10791} (2017) 1}.

\bibitem{2018ApJ...854L..32P}
S.~{Paiano}, R.~{Falomo}, A.~{Treves} and R.~{Scarpa}, \emph{{The Redshift of
  the BL Lac Object TXS 0506+056}},
  \href{https://doi.org/10.3847/2041-8213/aaad5e}{\emph{\apjl} {\bfseries 854}
  (2018) L32} [\href{https://arxiv.org/abs/1802.01939}{{\ttfamily
  1802.01939}}].

\bibitem{2017ATel10817....1M}
R.~{Mirzoyan}, \emph{{First-time detection of VHE gamma rays by MAGIC from a
  direction consistent with the recent EHE neutrino event IceCube-170922A}},
  {\emph{The Astronomer's Telegram} {\bfseries 10817} (2017) 1}.

\end{thebibliography}\endgroup

\clearpage
\begin{appendices}
\section{Questions and Answers}
\begin{itemize}
\item{\bf DENYS MALISHEV QUESTION:} What are current plans of MAGIC in CTA era? 

\noindent ANSWER:  MAGIC now is working very well  and will be operating during the next $\sim$ 5 years, when it is expected that the   four Large Size Telescopes (LST) of CTA-N will be available.  Then, there are different options  for the future use of the MAGIC telescopes although are under discussion and no decision has been taken yet. \end{itemize}
\end{appendices}

\end{document}